\documentclass[12pt]{article}
\usepackage[T1]{fontenc}
\usepackage{lmodern}
\usepackage[utf8]{inputenc}
\usepackage[english]{babel}
\usepackage[margin=1in]{geometry}

\usepackage{graphicx}
\usepackage{amsmath}
\usepackage{amssymb}
\usepackage{setspace}
\usepackage{multirow}
\usepackage{float}
\usepackage[super,sort&compress]{natbib}
\usepackage{natmove}
\usepackage{titling}
\usepackage[none]{hyphenat}
\usepackage{booktabs}
\usepackage{makecell}
\usepackage{arydshln}
\usepackage{siunitx}
\usepackage{amsmath}
\usepackage{url}
\usepackage{graphicx}
\usepackage{placeins}

\allowdisplaybreaks[4]

\usepackage{multirow}
\usepackage{threeparttable}

\usepackage{newtxtext}
\usepackage{newtxmath}

\newcommand{\mycite}[1]{\cite{#1}}
\setcitestyle{citesep={,}}

\usepackage{titling}

\pretitle{\LARGE\bfseries\centering}
\posttitle{\par\vspace{0.5em}}

\preauthor{\fontsize{12.5}{18}\selectfont\centering}
\postauthor{\par}

\predate{}
\postdate{\par\vspace{0em}}

\title{\begin{Large}\textbf{Scalable Matrix-Free Algorithms for Variance Estimation of Survey-Weighted Survival Curves}\end{Large}
\\
\vspace{20pt}
}

\author{
    {\fontsize{15}{18}\selectfont Jingchen Liang, MPH}\\[20pt]
    \makebox[\textwidth][c]{%
        \parbox{0.94\textwidth}{%
            \fontsize{14}{12.5}\selectfont
            \raggedright
            Yale School of Public Health,
            New Haven, Connecticut, USA\\[4pt]
        }%
    }
}

\date{}

\begin{document}

\raggedright
\setlength{\parindent}{1.5em}

\maketitle

\vspace{1.5cm}

\noindent\hspace*{0.03\textwidth}%
\begin{minipage}{0.94\textwidth}

\fontsize{14}{14}\selectfont

\textbf{Corresponding Author}

Jingchen Liang, MPH

Yale School of Public Health, New Haven, Connecticut 06511, USA

Email: jingchen.liang@aya.yale.edu

\end{minipage}

\newpage




\begingroup
\setlength{\parindent}{0pt}
\setlength{\parskip}{0.5em}

\vspace{0.2cm}

\begin{center}
\textbf{Abstract}
\end{center}

\vspace{-0.4cm}

Variance estimation for survey-weighted survival curves using existing statistical software can require substantial time and memory, particularly when the number of distinct event times is large. We developed two matrix-free algorithms that compute the conventional Taylor linearization variance estimator without constructing the full event-time covariance matrix: one accumulates individual-level linearization contributions, while the other directly updates aggregated contributions at the primary sampling unit (PSU) level. We evaluated numerical accuracy, runtime, and peak memory usage through an application to the National Health and Aging Trends Study (NHATS) and simulation experiments. Both approaches yielded survival and variance estimates identical to the R implementation. In the 7-year follow-up analysis of 7,608 NHATS participants with 2,879 deaths, computation time decreased from approximately 26 minutes to less than 3 seconds with the individual-level approach and less than 1 second with the PSU-level approach. Simulations with sample sizes of 1,000 to 10,000 showed similar results, with more pronounced computational advantages as sample size and the number of distinct event times increased. In more demanding settings, the existing implementation failed due to memory exhaustion, whereas both proposed approaches completed successfully. Matrix-free computation can improve the efficiency and scalability of design-based variance estimation while preserving numerical accuracy.

\par\medskip

\endgroup



\vspace{12pt}

\section{Introduction}

Survival curves are widely used in epidemiologic, biomedical, and population health research to describe time-to-event outcomes. Among nonparametric approaches, the Kaplan-Meier estimator is one of the most common methods for estimating the survival function without specifying a parametric distribution for survival times \mycite{kaplan1958nonparametric}. Nonparametric survival estimation is also increasingly applied to complex survey data, which typically require consideration of sampling weights, stratification, and clustering \mycite{lumley2010}. In this context, population-level inference requires not only appropriately weighted survival estimates but also valid design-based variance estimates to quantify uncertainty.

However, commonly used statistical software such as SAS and R still have limitations for nonparametric survival estimation with complex survey data, including the lack of standard procedures and limited computational efficiency for obtaining accurate design-based variance estimates. In SAS, nonparametric survival estimation such as weighted Kaplan-Meier estimation can be obtained using the \texttt{PROC LIFETEST} procedure with \texttt{WEIGHT} and \texttt{STRATA} statements, but these statements are not equivalent to a complex survey design because the \texttt{WEIGHT} is implemented using inverse-probability weighting and the \texttt{STRATA} defines group levels rather than sampling strata \mycite{SAS_user_guide, berglund2011overview}. SAS also provides \texttt{PROC SURVEYPHREG} procedure for semiparametric Cox proportional hazards regression to survey data, but this procedure is model-based and does not provide nonparametric survey-weighted survival estimation \mycite{SAS_user_guide}.

The \texttt{survey} package in R contains the \texttt{svykm()} function, which can estimate survey-weighted survival curves \mycite{surveypackage2026}. By default, \texttt{svykm()} returns a weighted Kaplan--Meier estimator, but when \texttt{se = TRUE}, it produces a Nelson--Aalen-based survival estimator with standard errors obtained through Taylor linearization \mycite{williams1995product,lumley2004analysis,surveypackage2026}. The underlying variance calculation, however, can require substantial computational resources, especially for large samples or a large number of events. In particular, the package documentation notes that requesting standard errors with \texttt{se = TRUE} can be slow for moderate to large data, with memory requirements proportional to the sample size multiplied by the square of the number of events \mycite{surveypackage2026}. Therefore, design-based variance estimation can be considerably more time-consuming or even infeasible as the number of distinct event times increases, which is common in studies with longer follow-up.

We encountered this limitation in our prior work on mortality ascertainment in the National Health and Aging Trends Study (NHATS) when estimating 10-year cumulative mortality curves with standard error bands for 7,608 participants in the 2011 analytic cohort \mycite{gill2025incomplete,gill2026losses}. This limitation became more apparent with longer follow-up as deaths accumulated and the number of distinct event times increased. In our computing environment, the computation exhausted the available memory while processing approximately 3,000 events and therefore could not be completed.


This experience motivated us to develop more efficient computational approaches that avoid constructing the full covariance matrix while retaining the same design-based variance estimator. We propose two scalable matrix-free algorithms based on cumulative Taylor linearization: an individual-level approach and a primary sampling unit (PSU)-level approach.

\section{Methods}
\label{sec:methods}

\subsection{Background: Survey-weighted survival estimation and implementation in R}

Consider a finite population and suppose that a probability sample $s$ of size $n$ is drawn from this population under a complex sampling design. For sampled individual $i$, let $w_i$ denote the survey weight, $T_i$ the event time, $C_i$ the right-censoring time, $X_i=\min(T_i,C_i)$ the observed follow-up time, and $\delta_i=I(T_i\leq C_i)$ the event indicator. We consider a stratified clustered design with strata $h=1,\ldots,H$ and clusters $p=1,\ldots,P_h$ within stratum $h$. In many complex survey designs, these clusters correspond to primary sampling units (PSUs).

Let $t_1<\cdots<t_m$ denote the $m$ distinct observed event times. For individual $i\in s$ and event time $t_j$, define
\begin{equation}
 d_{ij}=I(X_i=t_j,\delta_i=1), 
 \qquad
 y_{ij}=I(X_i\geq t_j),
 \label{equation_1}
\end{equation}
where $d_{ij}$ is the event indicator for subject $i$ at $t_j$, and $y_{ij}$ is the corresponding at-risk indicator at $t_j$.

At each event time $t_j$, the survey-weighted event and risk-set totals, as well as the corresponding estimated hazard increment, are defined as
\begin{equation}
 \widehat{D}_j=\sum_{i\in s}w_i d_{ij},
 \qquad
 \widehat{Y}_j=\sum_{i\in s}w_i y_{ij}.
 \qquad 
 \widehat h_j=\frac{\widehat{D}_j}{\widehat{Y}_j}.
 \label{equation_2}
\end{equation}
For an arbitrary time $t$, the cumulative hazard and survival function are
\begin{equation}
 \widehat H(t)=\sum_{j:t_j\leq t}\widehat h_j, 
 \qquad
 \widehat S(t)=\exp(-\sum_{j:t_j\leq t}\widehat h_j).
 \label{equation_3}
\end{equation}
This corresponds to the survey-weighted Nelson--Aalen cumulative hazard estimator obtained by \texttt{svykm()} in R when design-based standard errors are requested, with the survival curve given by $\widehat S(t)=\exp(-\widehat H(t))$\mycite{surveypackage2026}.

The design-based variance calculation in \texttt{svykm()} starts with a joint calculation of the weighted event and risk-set totals using \texttt{svytotal} \mycite{surveypackage2026}. For exposition, we express the design-based variance calculation at the distinct-event-time level. Let $\widehat{\boldsymbol q} = (\widehat D_1,\ldots,\widehat D_m,\widehat Y_1,\ldots,\widehat Y_m)^{\mathsf T}$ denote the joint vector of survey-weighted event and risk-set totals, and $\widehat{\boldsymbol\Sigma}_{q}=\widehat{\operatorname{Var}} (\widehat{\boldsymbol q})$ denote the corresponding $2m \times 2m$ design-based covariance matrix. The covariance matrix can then be partitioned into four blocks with three distinct covariance components: covariances among event totals, covariances among risk-set totals, and covariances between event and risk-set totals:
\begin{equation}
 \widehat {\boldsymbol\Sigma}_{q}= \begin{pmatrix} 
 \widehat{\boldsymbol\Sigma}_{DD} & \widehat{\boldsymbol\Sigma}_{DY} \\ 
 \widehat{\boldsymbol\Sigma}_{YD} & \widehat{\boldsymbol\Sigma}_{YY} 
 \end{pmatrix}
 \label{equation_4}
\end{equation}
Let $g(D,Y)=\dfrac{D}{Y}$. Then, at each time $t_j$, $\widehat h_j = g(\widehat{D}_j,\widehat{Y}_j)$. Using the first-order Taylor expansion, the delta method gives the following variance approximation for the estimated hazard increment at $t_j$:
\begin{equation}
 \begin{aligned}
 \widehat{\operatorname{Var}} (\widehat h_j) 
 &\approx 
    \left(\frac{\partial g}{\partial D}\right)^2 \widehat{\operatorname{Var}} (\widehat{D}_j) + 
    \left(\frac{\partial g}{\partial Y}\right)^2 \widehat{\operatorname{Var}} (\widehat{Y}_j) + 
    2\left(\frac{\partial g}{\partial D}\right) \left(\frac{\partial g}{\partial Y}\right) \widehat{\operatorname{Cov}}(\widehat{D}_j,\widehat{Y}_j) \\
 &= 
    \frac{1}{\widehat{Y}_j^2} \widehat{\operatorname{Var}} (\widehat{D}_j) +
    \frac{\widehat{D}_j^2}{\widehat{Y}_j^4} \widehat{\operatorname{Var}} (\widehat{Y}_j) -
    \frac{2 \widehat{D}_j}{\widehat{Y}_j^3} \widehat{\operatorname{Cov}}(\widehat{D}_j,\widehat{Y}_j)
 \end{aligned}
 \label{equation_5}
\end{equation}
Let $\widehat{\boldsymbol h} = (\widehat h_1,\ldots,\widehat h_m)^{\mathsf T}$ denote the vector of estimated hazard increments across all distinct event times. Let $\boldsymbol J$ denote the $m \times 2m$ Jacobian matrix for the mapping from $\widehat{\boldsymbol q}\in\mathbb{R}^{2m}$ to $\widehat{\boldsymbol h}\in\mathbb{R}^{m}$. Then,
\begin{equation}
 \begin{aligned}
 \boldsymbol J 
 &= \left(\operatorname{diag} \left(\frac{1}{\widehat Y_1},\ldots,\frac{1}{\widehat Y_m}\right) \quad 
    \operatorname{diag} \left(-\frac{\widehat D_1}{\widehat Y_1^2},\ldots,-\frac{\widehat D_m}{\widehat Y_m^2}\right)\right).
 \end{aligned}
 \label{equation_6}
\end{equation}
Using the multivariate delta method, the design-based covariance matrix of $\widehat{\boldsymbol h}$ can be approximated by
\begin{equation}
 \widehat {\boldsymbol\Sigma}_{h} \approx \boldsymbol J \widehat {\boldsymbol\Sigma}_{q} \boldsymbol J^{\mathsf T}.
 \label{equation_7}
\end{equation}
Equivalently, under the Jacobian transformation $\boldsymbol J \widehat {\boldsymbol\Sigma}_{q} \boldsymbol J^{\mathsf T}$, the $(j,k)$ element of $\widehat {\boldsymbol\Sigma}_{h}$ can be expressed based on the four blocks of $\widehat {\boldsymbol\Sigma}_{q}$ described in Equation~\eqref{equation_4} as
\begin{equation}
 \widehat{\operatorname{Cov}} (\widehat h_j,\widehat h_k) 
 \approx 
    \frac{\widehat{\operatorname{Cov}}(\widehat{D}_j,\widehat D_k)}{\widehat{Y}_j \widehat Y_k} + 
    \frac{\widehat{D}_j \widehat D_k}{\widehat{Y}_j^2 \widehat Y_k^2}\widehat{\operatorname{Cov}}(\widehat{Y}_j,\widehat Y_k) -
    \frac{\widehat{D}_j}{\widehat{Y}_j^2 \widehat Y_k}\widehat{\operatorname{Cov}}(\widehat{Y}_j,\widehat D_k) - 
    \frac{\widehat D_k}{\widehat{Y}_j \widehat Y_k^2}\widehat{\operatorname{Cov}}(\widehat{D}_j,\widehat Y_k).
 \label{equation_8}
\end{equation}
When $j = k$, Equation~\eqref{equation_8} reduces to the single-time variance approximation in Equation~\eqref{equation_5}. According to Equation~\eqref{equation_3}, the cumulative hazard at the $r$th event time $t_r$ is $\displaystyle \widehat H(t_r)=\sum_{j=1}^{r}\widehat h_j$, and its variance is 
\begin{equation}
 \widehat{\operatorname{Var}} \left[\widehat H(t_r)\right] 
 =\widehat{\operatorname{Var}} \left[\sum_{j=1}^{r}\widehat h_j\right]
 =\sum_{j=1}^{r}\sum_{k=1}^{r}\widehat {\operatorname{Cov}} (\widehat h_j,\widehat h_k).
 \label{equation_9}
\end{equation}
Because $\widehat{\operatorname{Var}} \left[{\operatorname{log}}(\widehat S(t_r))\right]=\widehat{\operatorname{Var}} \left[\widehat H(t_r)\right]$, the \texttt{svykm()} function returns the survival estimate $\exp(-\widehat H(t_r))$ and stores the cumulative hazard variance as the variance of the log survival estimate\mycite{surveypackage2026}.

A key computational limitation of the current \texttt{svykm()} function is the explicit construction of the full design-based covariance matrix of the survey-weighted event and the risk-set totals. Also, tied events are not first aggregated, with each observed event contributing a separate event column and a risk-set column. Let $d$ denote the total number of observed events. The corresponding joint covariance matrix has dimension $2d \times 2d$, where $d \geq m$, so the current implementation operates on a larger covariance matrix than the one based on distinct event times when tied events are present. In addition, the delta-method transformation also requires operating on large covariance matrices to obtain a $d \times d$ covariance matrix of the hazard increments. Thus, the computational burden becomes increasingly substantial as the number of observed events increases.

\subsection{Individual-level matrix-free linearization}

The conventional covariance-based approach described in Section~2.1 first estimates the design-based variances and covariances of the linearized hazard increments across event times and then sums these variance and covariance terms to obtain the variance of the cumulative hazard. The same first-order Taylor-linearized variance can also be obtained by reversing these two steps: first summing the linearized contributions across event times within each sampled individual, and then estimating the design-based variance from the resulting cumulative contributions.

Given $\widehat{h}_j = g(\widehat{D}_j,\widehat{Y}_j)$, a first-order Taylor expansion with respect to small perturbations in the weighted event total $\widehat{D}_j$ and weighted risk-set total $\widehat{Y}_j$ gives

\[
 \begin{aligned}
g(\widehat{D}_j+\Delta \widehat{D}_j,\widehat{Y}_j+\Delta \widehat{Y}_j) 
  &\approx
    g(\widehat{D}_j,\widehat{Y}_j)
    + \frac{\partial g}{\partial \widehat{D}_j} \Delta\widehat{D}_j
    + \frac{\partial g}{\partial \widehat{Y}_j} \Delta\widehat{Y}_j \\
  &=
    g(\widehat{D}_j,\widehat{Y}_j)
    + \frac{1}{\widehat{Y}_j}\Delta \widehat{D}_j
    - \frac{\widehat{D}_j}{\widehat{Y}_j^2} \Delta \widehat{Y}_j.
 \end{aligned}
\]
The corresponding perturbation in the estimated hazard increment is
\begin{equation}
 \begin{aligned}
\Delta \widehat h_j
  \approx
    \frac{1}{\widehat{Y}_j}\Delta \widehat{D}_j 
    -
    \frac{\widehat{D}_j}{\widehat{Y}_j^2} \Delta \widehat{Y}_j.
 \end{aligned}
 \label{equation_10}
\end{equation}
We next consider how such perturbations are induced by a small change in the weight of subject $i$. Specifically, suppose $w_i \rightarrow w_i+\varepsilon$, while holding the weights of all other sampled individuals fixed. This gives $\Delta \widehat D_j = \varepsilon d_{ij}$ and $\Delta \widehat Y_j = \varepsilon y_{ij}$. Then,
\[
\begin{aligned}
\Delta \widehat h_j
  \approx
    \frac{1}{\widehat{Y}_j}\Delta \widehat{D}_j - \frac{\widehat{D}_j}{\widehat{Y}_j^2} \Delta \widehat{Y}_j 
  = 
    \frac{1}{\widehat{Y}_j}\varepsilon d_{ij} - \frac{\widehat{D}_j}{\widehat{Y}_j^2} \varepsilon y_{ij} 
  = 
    \varepsilon \left(\frac{d_{ij}}{\widehat{Y}_j} - \frac{\widehat{D}_j}{\widehat{Y}_j^2} y_{ij}\right).
\end{aligned}
\]
Thus, we define the corresponding individual-level linearized contribution at event time $t_j$ as

\begin{equation}
\ell_{ij} = \frac{d_{ij}}{\widehat Y_j} - \frac{\widehat D_j}{\widehat Y_j^2}y_{ij},
 \label{equation_11}
\end{equation}
where $\ell_{ij}$ represents the first-order contribution of sampled
individual $i$ to the sampling variation of the estimated hazard
increment $\widehat h_j$.

Because the cumulative hazard at $t_r$ is the sum of the first $r$
hazard increments, the corresponding cumulative linearized
contribution for individual $i$ is
\[
L_i(t_r) = \sum_{j=1}^{r}\ell_{ij}.
\]
Equivalently, $L_i(t_r)$ can be updated recursively as 
\[
L_i(t_r) =L_i(t_{r-1})+\ell_{ir}, \quad \text{with } L_i(t_0)=0.
\]
Thus, the recursive algorithm only needs to retain the cumulative linearized contribution $L_i(t_r)$ for each sampled individual, updating it at each distinct event time. We then apply \texttt{svytotal} to the cumulative linearized contributions $L_i(t_r)$ for all $n$ sampled individuals using the original survey design. The corresponding survey-weighted total can be written as
\begin{equation}
\sum_{i\in s}w_iL_i(t_r)
=
\sum_{i\in s}w_i\sum_{j=1}^{r}\ell_{ij}
=
\sum_{j=1}^{r}\sum_{i\in s}w_i\ell_{ij}.
\label{equation_cumulative_linearized_total}
\end{equation}
Then, treating the estimated linearization coefficients as fixed, the design-based variance of the survey-weighted total can be expressed as
\begin{align}
\widehat{\operatorname{Var}} \left[\sum_{i\in s}w_iL_i(t_r)\right]
&=
\widehat{\operatorname{Var}} \left[\sum_{j=1}^{r}\sum_{i\in s}w_i\ell_{ij}\right] \notag\\
&=
\sum_{j=1}^{r}\sum_{k=1}^{r} \widehat{\operatorname{Cov}}
\left(
\sum_{i\in s}w_i\ell_{ij},
\sum_{i\in s}w_i\ell_{ik}
\right) \notag\\
&=
\sum_{j=1}^{r}\sum_{k=1}^{r} \widehat{\operatorname{Cov}}
\left(
\sum_{i\in s}w_i \left(\frac{d_{ij}}{\widehat Y_j} - \frac{\widehat D_j}{\widehat Y_j^2}y_{ij}\right), \
\sum_{i\in s}w_i \left(\frac{d_{ik}}{\widehat Y_k} - \frac{\widehat D_k}{\widehat Y_k^2}y_{ik}\right)
\right) \notag\\
&=
\sum_{j=1}^{r}\sum_{k=1}^{r} \widehat{\operatorname{Cov}}
\left(
\frac{1}{\widehat Y_j}\sum_{i\in s}w_id_{ij} - \frac{\widehat D_j}{\widehat Y_j^2} \sum_{i\in s}w_iy_{ij}, \
\frac{1}{\widehat Y_k}\sum_{i\in s}w_id_{ik} - \frac{\widehat D_k}{\widehat Y_k^2} \sum_{i\in s}w_iy_{ik}
\right) \notag\\
&=
\sum_{j=1}^{r}\sum_{k=1}^{r}
\Biggl(
\frac{\widehat{\operatorname{Cov}}(\widehat{D}_j,\widehat D_k)}{\widehat{Y}_j \widehat Y_k}
- \frac{\widehat D_k}{\widehat{Y}_j \widehat Y_k^2}\widehat{\operatorname{Cov}}(\widehat{D}_j,\widehat Y_k) \notag\\
&\hspace{4.5em}
- \frac{\widehat{D}_j}{\widehat{Y}_j^2 \widehat Y_k}\widehat{\operatorname{Cov}}(\widehat{Y}_j,\widehat D_k)
+ \frac{\widehat{D}_j \widehat D_k}{\widehat{Y}_j^2 \widehat Y_k^2}\widehat{\operatorname{Cov}}(\widehat{Y}_j,\widehat Y_k)
\Biggr) \notag\\
&=
\sum_{j=1}^{r}\sum_{k=1}^{r}
\widehat{\operatorname{Cov}}(\widehat h_j,\widehat h_k) \qquad  \text{identical to Equation~\eqref{equation_8}} 
\notag\\
&=
\widehat{\operatorname{Var}}\left[\widehat H(t_r)\right].
\label{eq:variance}
\end{align}
The derivation above shows that, for each pair of event times $t_j$ and $t_k$, the four covariance terms reproduce the first-order Taylor-linearized covariance expression in Equation~\eqref{equation_8}. Summing these covariance estimates yields the same estimated variance of the cumulative hazard $\widehat H(t_r)$, as given in Equation~\eqref{equation_9}. Thus, the first-order Taylor-linearized design variance obtained from the individual-level formulation is equal to that obtained using the approach described in Section~2.1.

Specifically, unlike the approach in Section~2.1, which computes cross-time covariances before aggregation, the matrix-free formulation first accumulates the linearized contributions across event times within each sampled individual and then estimates the design variance of the resulting cumulative variable. This order of computation avoids constructing or storing the full event-time covariance matrix. Because the cumulative linearized contributions form a single survey variable, the estimated cumulative-hazard variance is the only entry in the resulting $1\times1$ covariance matrix.

\subsection{PSU-level matrix-free linearization}

To further improve computational efficiency, we can aggregate the linearized contributions at the PSU level. Similar to the survey-weighted total over all sampled individuals in Equation~\eqref{equation_cumulative_linearized_total} of
Section~2.2, we define the cumulative contribution for PSU $p$ within stratum $h$ as
\begin{equation}
\begin{aligned}
G_{hp}(t_r)
&=
\sum_{i\in s_{hp}}w_iL_i(t_r) \\
&=
\sum_{i\in s_{hp}}w_i\sum_{j=1}^{r}\ell_{ij} \\
&=
\sum_{j=1}^{r}\sum_{i\in s_{hp}}w_i\ell_{ij} \\
&=
\sum_{j=1}^{r}
\left (\frac{1}{\widehat Y_j}\sum_{i\in s_{hp}}w_id_{ij} - \frac{\widehat D_j}{\widehat Y_j^2} \sum_{i\in s_{hp}}w_iy_{ij}
\right) \\
&=
\sum_{j=1}^{r}
\left(
\frac{\widehat D_{hp,j}}{\widehat Y_j}
-
\frac{\widehat D_j}{\widehat Y_j^2}
\widehat Y_{hp,j}
\right).
\end{aligned}
\label{equation_psu_cumulative}
\end{equation}
where $s_{hp}$ denotes the set of sampled individuals in PSU $p$ within stratum $h$, and $\widehat D_{hp,j}$ and $\widehat Y_{hp,j}$ denote the corresponding weighted event total and weighted risk-set total at time $t_j$, respectively. Similarly, the cumulative PSU-level contribution can be updated recursively as
\[
G_{hp}(t_r) = G_{hp}(t_{r-1}) + \frac{\widehat D_{hp,r}}{\widehat Y_r} - \frac{\widehat D_r}{\widehat Y_r^2}\widehat Y_{hp,r},
\quad \text{with } G_{hp}(t_0)=0.
\label{equation_psu_recursive}
\]
Let $\bar G_h(t_r) = \frac{1}{P_h} \sum_{p=1}^{P_h}G_{hp}(t_r)$ denote the mean cumulative linearized contribution across PSUs within stratum $h$ at time $t_r$. The total variance estimate is then obtained by summing the stratum-specific variance contributions calculated from the cumulative PSU-level contributions:
\begin{equation}
\widehat{\operatorname{Var}}_{\mathrm{PSU}}(t_r)
= \sum_{h=1}^{H} \mathrm{FPC}_h \frac{P_h}{P_h-1} \sum_{p=1}^{P_h} \left(G_{hp}(t_r)-\bar G_h(t_r)\right)^2.
\label{equation_psu_variance}
\end{equation}
where $\mathrm{FPC}_h$ is the finite population correction factor for stratum $h$. When $\mathrm{FPC}_h=1$ for all strata, the variance estimator corresponds to a first-stage with-replacement approximation, and no finite population correction is applied \mycite{cochran1977sampling,lumley2004analysis,chromy2005statistical}. Since $G_{hp}(t_r)$ is exactly the weighted sum of the individual-level cumulative linearized contributions within each PSU, this calculation yields the same design-based variance estimate as the individual-level approach in Section 2.2, under the same survey-design specifications.

In the individual-level approach of Section 2.2, cumulative linearized contributions are updated for all sampled individuals at each event time and subsequently aggregated by PSU for variance estimation. The PSU-level formulation computes the increments directly from the weighted PSU-level event and risk-set totals and updates $G_{hp}(t_r)$ recursively. This avoids explicitly constructing individual-level linearized increments and repeatedly aggregating individual cumulative contributions, while reducing the number of stored cumulative contributions from the total number of sampled individuals to the total number of sampled PSUs.

\section{Computational evaluation}
We compared \texttt{svykm()}, the individual-level matrix-free implementation, and the PSU-level matrix-free implementation in terms of their survival estimates, variance estimates, runtime, and peak memory usage. Analyses were conducted in R, using version 4.4.1 for the NHATS analyses and version 4.6.1 for the simulation studies.

\subsection{Application to NHATS}
In our previous mortality ascertainment analyses using NHATS data \mycite{gill2025incomplete,gill2026losses}, we aimed to estimate 10-year cumulative mortality and corresponding 95\% confidence intervals for the 2011 cohort. This required estimating survey-weighted survival curves and their standard errors, for which we initially used \texttt{svykm()} with \texttt{se = TRUE}. However, as the number of observed events increased over follow-up, the memory requirements of the full-covariance approach made direct variance estimation over the complete 10-year follow-up period computationally impractical. We therefore used a blockwise covariance extension procedure based on the structure in Equation~\eqref{equation_4}. Covariance entries for the earlier follow-up were first calculated and retained. We then calculated the additional covariances among event and risk-set totals at later event times, together with the cross-covariances between the earlier and later totals. The retained and newly calculated entries were combined to form the full covariance matrix, which was used to calculate the variance estimates over the complete follow-up period. For the current comparison, this procedure provided the reference estimates at 8 and 10 years of follow-up.


We evaluated the three approaches using the 2011 NHATS cohort sample ($N=7{,}608$) at 3, 5, 7, 8, and 10 years of follow-up. The NHATS analysis used 56 strata, with 2 clusters (PSUs) per stratum for variance estimation \mycite{freedmanFAQ,montaquilaR1Wgts}. No finite population correction was used, with $\mathrm{FPC}_h$ set to 1 for all strata. Runtime and peak RAM usage were evaluated over 30 and 5 repetitions, respectively. We also calculated the maximum absolute differences in survival estimates and log-survival variance estimates between each matrix-free implementation and the reference estimates obtained using \texttt{svykm()} or the blockwise extension procedure within each follow-up period. All computations were performed within the secure computing environment used for restricted NHATS data \mycite{nhatsCMSaccess}. The NHATS protocol was approved by the Johns Hopkins University Institutional Review Board, and participants provided written informed consent. Use of NHATS data for this analysis was approved by the Yale University Institutional Review Board.

\subsection{Simulation studies}
A simulation study was conducted to evaluate computational performance across different sample sizes and numbers of events. Twelve simulated datasets were generated using six sample sizes $n \in \{1{,}000, 2{,}000, 4{,}000, 6{,}000, 8{,}000, 10{,}000\}$ and two observed event proportions $D/n \in \{0.25, 0.50\}$, where $D$ denotes the number of observed deaths. In each dataset, individuals were randomly assigned to one of 20 strata and one of 4 clusters (PSUs) within each stratum. Event times were generated from a Weibull proportional hazards model incorporating stratum-level and PSU-level random effects and an individual-level covariate \mycite{sun2006statistical}. The shape parameter was set to 2 to allow the hazard to increase over time. Individual-specific censoring times were independently generated from an exponential distribution with rate 0.05, and survey weights were independently drawn from a uniform distribution between 1 and 5,000. For each dataset, a single administrative censoring time was selected to obtain exactly $D$ observed deaths. Each individual's observed follow-up time was defined as the minimum of their event time, independent censoring time, and the administrative censoring time. For each simulated dataset, we evaluated the three computational approaches using the same procedures as in Section~3.1 to compare estimation accuracy, runtime, and peak memory usage. Each run was limited to 30 minutes, with runs exceeding this limit recorded as timeouts.

\section{Results}
\textbf{Table~\ref{tab:NHATS}} compares the computational performance of the three approaches using the NHATS sample. Survival estimates and log-survival variance estimates closely matched the reference results, with maximum absolute differences of approximately $10^{-16}$ and $10^{-19}$, respectively. The median runtime of \texttt{svykm()} increased substantially with the number of events. Both matrix-free implementations greatly reduced computation time: the individual-level implementation completed the calculations within a few seconds, whereas the PSU-level implementation further reduced median runtime to less than one second at each follow-up time. At 8 and 10 years of follow-up, direct computation using \texttt{svykm()} failed because of memory exhaustion associated with the full-covariance calculation. In contrast, the individual-level and PSU-level matrix-free implementations completed these analyses with median runtimes of less than 4 seconds and less than 1 second, respectively. Given that both implementations avoid constructing the full event-time covariance matrix, they also considerably reduced peak RAM usage. These results suggest that the runtime advantages of the matrix-free implementations are maintained and may become more pronounced with longer follow-up.

\begin{table}[htbp]
\centering
\caption{Computational performance comparison in the NHATS sample ($N=7{,}608$).}
\label{tab:NHATS}
\renewcommand{\arraystretch}{1.5}

\makebox[\textwidth][c]{%
\begin{threeparttable}
\begin{tabular}{lccccccc}
\toprule
\makecell{Follow-\\up} & Method & \makecell{Total\\Events} & \makecell{Max\\$|\Delta S|$} & \makecell{Max\\$|\Delta V|$} & \makecell{Median\\Runtime\\(IQR), sec} & \makecell{Median\\Peak RAM\\MiB} & \makecell{Status}  \\
\midrule
\multirow{3}{*}{3 years}
 & svykm() & \multirow{3}{*}{$1{,}218$} & -- & -- & 345.26 (78.13) & $18{,}872.70$ & $\checkmark$ \\
 & Individual-level & & $<10^{-15}$ & $<10^{-20}$ & 1 (0.02) & 41.90 & $\checkmark$ \\
 & PSU-level & & $<10^{-15}$ & $<10^{-20}$ & 0.15 (0.02) & 12.30 & $\checkmark$  \\
\hdashline
\multirow{3}{*}{5 years}
 & svykm() & \multirow{3}{*}{$2{,}048$} & -- & -- & 768.37 (124.73) & $50{,}440.90$ & $\checkmark$   \\
 & Individual-level &  & $<10^{-15}$ & $<10^{-18}$ & 1.81 (0.03) & 56.4 & $\checkmark$  \\
 & PSU-level & & $<10^{-15}$ & $<10^{-18}$ & 0.22 (0.01) & 16.2 & $\checkmark$  \\
\hdashline
\multirow{3}{*}{7 years}
 & svykm() & \multirow{3}{*}{$2{,}879$} & -- & -- & $1{,}547.96$ ($256.01$) & $81{,}056.60$ & $\checkmark$ \\
 & Individual-level & & $<10^{-15}$ & $<10^{-18}$ & 2.36 (0.04) & 61.00 & $\checkmark$  \\
 & PSU-level & & $<10^{-15}$ & $<10^{-18}$ & 0.29 (0.02) & 23.40 & $\checkmark$  \\
\hdashline
\multirow{3}{*}{8 years}
 & svykm() & \multirow{3}{*}{$3{,}268$} & -- & -- & N/A & N/A & $\times$ \\
 & Individual-level & & $<10^{-15}$ & $<10^{-18}$ & 2.67 (0.03) & 63.90 & $\checkmark$ \\
 & PSU-level & & $<10^{-15}$ & $<10^{-18}$ & 0.32 (0.02) & 26.00 & $\checkmark$ \\
\hdashline
\multirow{3}{*}{10 years}
 & svykm() & \multirow{3}{*}{$4{,}045$} & -- & -- & N/A & N/A & $\times$ \\
 & Individual-level & & $<10^{-15}$ & $<10^{-18}$ & 3.38 (0.03) & 65.80 & $\checkmark$  \\
 & PSU-level & & $<10^{-15}$ & $<10^{-18}$ & 0.39 (0.01) & 31.8 & $\checkmark$  \\

\bottomrule
\end{tabular}

\begin{tablenotes}[flushleft]
\footnotesize
\item \textit{Note:} Max $|\Delta S|$ and Max $|\Delta V|$ denote the maximum absolute differences in survival estimates and log-survival variance estimates, respectively, relative to the reference estimates; these differences are reported as upper bounds in powers of 10. Reference estimates were obtained directly from \texttt{svykm()} for the 3-, 5-, and 7-year analyses and using the blockwise covariance extension procedure described in Section~3.1 for the 8- and 10-year analyses. Runtime and peak RAM were summarized across repeated runs. Status: $\checkmark$ indicates successful completion, whereas $\times$ indicates that direct computation could not be completed because of memory exhaustion. Variances were estimated using a first-stage with-replacement approximation, with no finite population correction ($\mathrm{FPC}_h=1$ for all strata). All NHATS benchmarks reported in this table were performed within the secure computing environment used for restricted NHATS data.
\end{tablenotes}
\end{threeparttable}
}
\end{table}

\begin{figure}[htbp]
\centering
\caption{Median runtime and peak memory usage of the individual-level and PSU-level methods in simulation studies.}
\label{fig:simulation}
\begin{threeparttable}
    \includegraphics[width=\textwidth]{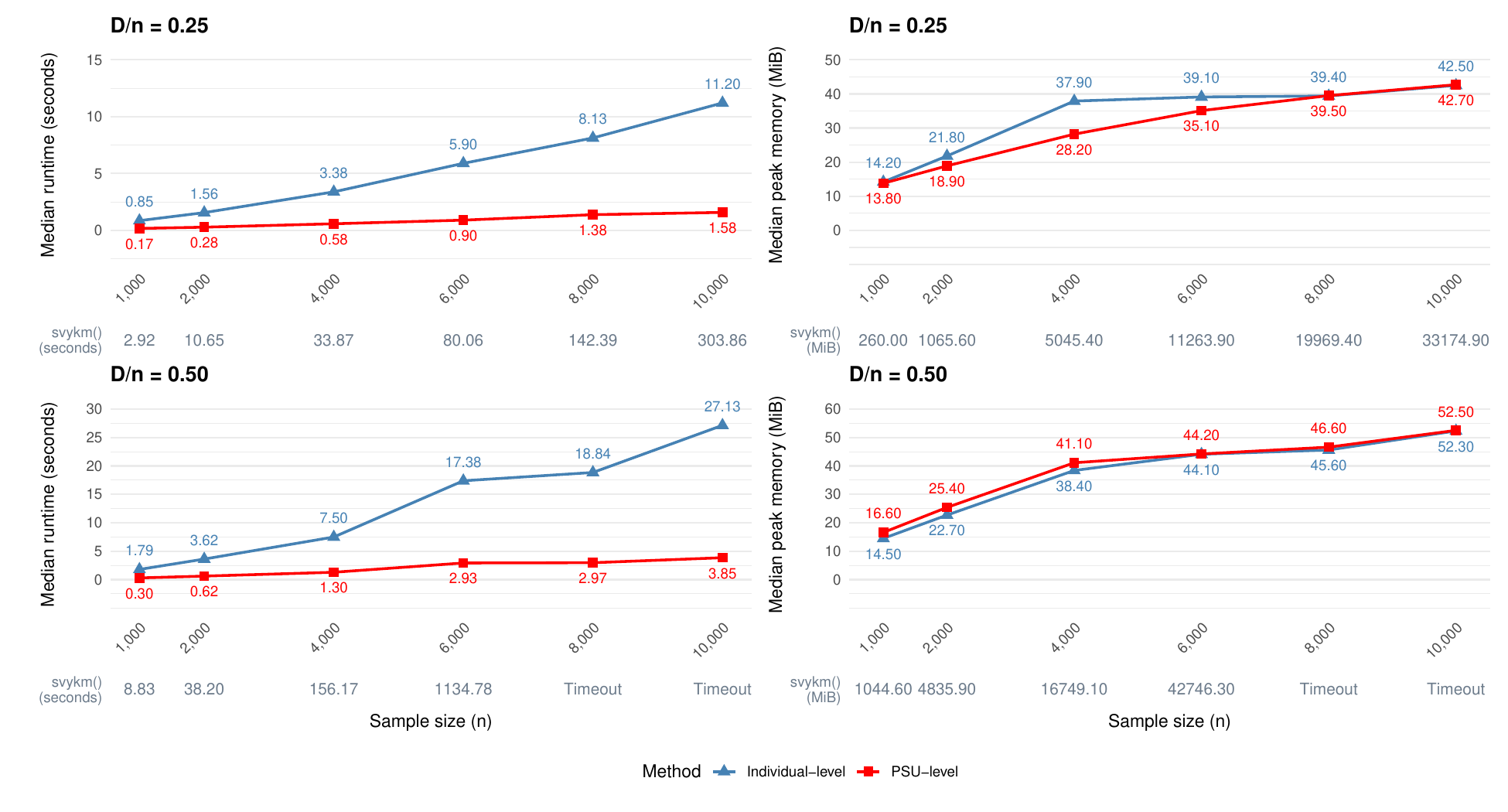}
    \begin{tablenotes}[flushleft]
    \footnotesize
    \item \textit{Note:} Median runtime and peak RAM were summarized across repeated runs. $D/n$ represents the proportion of observed deaths, where $D$ is the number of observed deaths and $n$ is the sample size. Results from \texttt{svykm()} are presented separately below each panel because of their substantially larger runtime and memory requirements. Timeout indicates that a run exceeded the 30-minute limit.
    \end{tablenotes}
\end{threeparttable}
\end{figure}

Similar results were observed in the simulation study (\textbf{Figure~\ref{fig:simulation}}). Both matrix-free implementations markedly reduced runtime and peak memory usage compared with \texttt{svykm()}. The PSU-level implementation was consistently faster than the individual-level implementation, with a greater runtime advantage as the sample size and number of events increased. For example, at $n = 10{,}000$ and $D/n = 0.50$, median runtime was 3.85 seconds for the PSU-level implementation compared with 27.13 seconds for the individual-level implementation, whereas \texttt{svykm()} exceeded the 30-minute time limit. In addition, peak memory usage was similar between the two matrix-free implementations. Unlike NHATS, where deaths were recorded by month and many individuals had the same recorded event time, the simulated event times were more granular, with no tied event times. Additional tests showed that rounding simulated event times to create ties resulted in a greater reduction in peak memory usage for the PSU-level implementation than for the individual-level implementation. These results suggest that the number of distinct event times has a larger effect on peak memory usage for the PSU-level implementation than for the individual-level implementation.

\FloatBarrier

\section{Discussion}
To address the computational burden of variance estimation for survey-weighted survival curves using existing statistical software, we developed two matrix-free algorithms, one at the individual level and the other at the PSU level, by reformulating the existing Taylor-linearized variance estimator to avoid constructing the full covariance matrix. Both proposed approaches provided accurate estimates while significantly reducing computation time and memory usage, making them particularly useful for large datasets with many observed events. 

In our previous NHATS research \mycite{gill2025incomplete,gill2026losses}, estimating 10-year cumulative mortality using \texttt{svykm()} became computationally infeasible at later follow-up years, requiring calculations over shorter follow-up periods followed by a manual blockwise covariance extension procedure. The newly developed matrix-free approaches eliminated the need for these manual calculations, allowing direct estimation over the full follow-up period. Beyond the NHATS application, simulation results also demonstrated considerable computational improvements across different sample sizes and numbers of observed events.


The computational advantage arises from the recursive updating of cumulative linearized contributions across event times, rather than from modifying the underlying variance estimator. This recursive procedure incorporates cross-time covariance information without explicitly calculating and storing individual covariance terms. Importantly, the proposed approaches preserve the original survey-weighted Nelson--Aalen cumulative hazard estimator and its design-based Taylor-linearized variance estimator, achieving computational improvements without introducing additional statistical assumptions or approximations. Such a recursive updating strategy is particularly advantageous for studies with many distinct event times, where the conventional implementation may require storing increasingly large covariance matrices. For example, in the 5-year follow-up analysis using NHATS data, 2,048 deaths were observed, and \texttt{svykm()} required the construction and manipulation of a $4{,}096 \times 4{,}096$ joint covariance matrix containing more than 16 million entries. In contrast, both proposed approaches avoided constructing such a high-dimensional matrix.

Compared with the individual-level implementation, the PSU-level implementation further reduced computation time by recursively updating PSU-level totals rather than maintaining individual-level cumulative contributions. Because the design-based variance estimator depends on variation in PSU totals within strata, individual-level cumulative contributions need not be maintained throughout the calculation. Furthermore, unlike the individual-level implementation, which relies on repeated calls to \texttt{svytotal()}, the PSU-level implementation calculates the variance directly from the updated PSU totals, avoiding the associated computational overhead. These advantages may be particularly relevant when the number of sampled PSUs is much smaller than the number of individuals. Nevertheless, the individual-level implementation remains a valuable alternative, as both approaches offer substantial computational improvements over the conventional implementation.

The memory advantage of the PSU-level implementation may be less pronounced when event times are more granular, resulting in a larger number of distinct event times. In practice, however, event times in many observational studies, including dates of death, are recorded at the day or month level and are therefore less granular than the event times initially generated on a continuous scale in our simulations. For example, NHATS obtains the month and year of death from proxy respondents through its Last Month of Life interview \mycite{nhatsLML}, whereas Medicare data from the Centers for Medicare \& Medicaid Services (CMS) record dates of death at the day level, with the last day of the month assigned when the day of death is unverified \mycite{resdac2022death}. As a result, multiple individuals may share the same recorded event time, resulting in more ties and potentially allowing the PSU-level implementation to achieve greater memory savings in real-world applications.

The proposed approaches are specifically designed for the Taylor-linearized variance estimator and may not be directly applicable to other variance estimation methods, such as balanced repeated replication (BRR), jackknife, and bootstrap. Future work could investigate efficient implementations for these alternative variance estimation methods. In addition, similar computational strategies could be extended to covariate-adjusted survival curves from survey-weighted Cox proportional hazards models \mycite{binder1992fitting}, where variance estimation can be computationally demanding. For example, the R \texttt{survey} package also documents long runtimes and high memory requirements for estimating standard errors of Cox model-based survival curves using \texttt{svycoxph()} \mycite{surveypackage2026}, while previous work and SAS technical guidance have highlighted challenges in obtaining survival predictions and valid design-based standard errors from survey-weighted Cox models using \texttt{PROC SURVEYPHREG} \mycite{jing2019survey,sasSurveyphregSupport}. These challenges highlight opportunities for further improvements in computational efficiency and software implementation for covariate-adjusted survival analysis with complex survey data.

\section{Conclusion}
The two proposed matrix-free approaches provide computationally efficient and scalable alternatives to conventional variance estimation for survey-weighted survival curves. By substantially reducing computation time and memory requirements while preserving the underlying Taylor-linearized variance estimator, these approaches may facilitate the analysis of large survey datasets with many distinct event times.


\section*{Acknowledgements}
I would like to express my sincere gratitude to Dr. Thomas M. Gill for his support of the NHATS research that motivated this work and to Dr. Brent Vander Wyk and Dr. Kendra Davis-Plourde for their helpful suggestions.

\section*{Funding}
This research was funded by the National Institute on Minority Health and Health Disparities, National Institutes of Health, grant number R01MD017298. The study was conducted at the Yale Claude D. Pepper Older Americans Independence Center (P30AG021342). NHATS is supported by U01AG032947.

\section*{Conflicts of Interest}
The author has no conflicts of interest.

\section*{Data Availability}

The NHATS restricted data used in this study are not publicly available. Researchers must obtain approval and sign a data use agreement to access these data.

\newpage

\begin{singlespace}



\bibliographystyle{SageV}
\bibliography{citations.bib}

@article{kaplan1958nonparametric,
  title={Nonparametric estimation from incomplete observations},
  author={Kaplan, Edward L and Meier, Paul},
  journal={Journal of the American statistical association},
  volume={53},
  number={282},
  pages={457--481},
  year={1958},
  publisher={Taylor \& Francis}
}

@article{williams1995product,
  title={Product-limit survival functions with correlated survival times},
  author={Williams, Rick L},
  journal={Lifetime data analysis},
  volume={1},
  number={2},
  pages={171--186},
  year={1995},
  publisher={Springer}
}

@article{lumley2004analysis,
  title={Analysis of complex survey samples},
  author={Lumley, Thomas},
  journal={Journal of statistical software},
  volume={9},
  pages={1--19},
  year={2004}
}

@Misc{surveypackage2026,
    author = {Thomas Lumley},
    year = {2026},
    title = {survey: analysis of complex survey samples},
    note = {R package version 4.5},
}

@Book{lumley2010,
year = {2010},
author = {Thomas Lumley},
title = {Complex Surveys: A Guide to Analysis Using R},
publisher = {John Wiley and Sons},
}

@book{SAS_user_guide,
  author    = {{SAS Institute Inc.}},
  title     = {{SAS/STAT} 15.4 User's Guide},
  year      = {2025},
  publisher = {SAS Institute Inc.},
  address   = {Cary, NC}
}

@article{berglund2011overview,
  title={An overview of survival analysis using complex sample data},
  author={Patricia A. Berglund},
  journal={Ann Arbor, MI: Institute for Social Research},
  year={2011}
}

@article{gill2025incomplete,
  title={Incomplete ascertainment of mortality in a nationally representative longitudinal study of community-living older Americans},
  author={Gill, Thomas M and Liang, Jingchen and Vander Wyk, Brent and Leo-Summers, Linda and Wang, Yi and Becher, Robert D and Davis-Plourde, Kendra},
  journal={The Journals of Gerontology, Series A: Biological Sciences and Medical Sciences},
  volume={80},
  number={7},
  pages={glaf014},
  year={2025},
  publisher={Oxford University Press US}
}

@article{gill2026losses,
  title={Losses to Follow-Up in a Nationally Representative Study of Community-Living Older Americans: Two Approaches to Censoring and Implications for Mortality Outcomes},
  author={Gill, Thomas M and Liang, Jingchen and Becher, Robert D and Davis-Plourde, Kendra},
  journal={Journal of the American Geriatrics Society},
  volume={74},
  number={1},
  pages={186--192},
  year={2026},
  publisher={Wiley Online Library}
}

@article{cochran1977sampling,
  title={Sampling techniques},
  author={Cochran, William G},
  journal={Johan Wiley \& Sons Inc},
  year={1977}
}

@incollection{chromy2005statistical,
  author    = {Chromy, James R. and Abeyasekera, Savitri},
  title     = {Statistical analysis of survey data},
  booktitle = {Household Sample Surveys in Developing
               and Transition Countries},
  series    = {Studies in Methods, Series F},
  number    = {96},
  chapter   = {19},
  pages     = {389--417},
  publisher = {United Nations},
  address   = {New York},
  year      = {2005},
  url       = {https://unstats.un.org/unsd/HHsurveys/pdf/Chapter_19.pdf}
}

@techreport{freedmanFAQ,
  author      = {Freedman, Vicki A. and Hu, Mengyao},
  title       = {Accounting for Sample Design in {NHATS} Analyses:
                 Frequently Asked Questions},
  institution = {Johns Hopkins Bloomberg School of Public Health},
  type        = {{NHATS} Technical Paper},
  number      = {55},
  year        = {2025},
  note        = {Available at \url{https://www.nhats.org/sites/default/files/public-documentation/NHATSTechnicalPaper55_09042025.pdf}}
}

@techreport{montaquilaR1Wgts,
  author      = {Montaquila, Jill and Freedman, Vicki A. and
                 Spillman, Brenda and Kasper, Judith D.},
  title       = {National Health and Aging Trends Study Development
                 of Round 1 Survey Weights, Updated February 2026},
  institution = {Johns Hopkins University School of Public Health},
  address     = {Baltimore},
  type        = {{NHATS} Technical Paper},
  number      = {2},
  year        = {2012},
  note        = {Available at \url{https://www.nhats.org/sites/default/files/public-documentation/NHATSTechnicalPaper2_Feb2026.pdf}}
}

@misc{nhatsCMSaccess,
  author       = {{National Health and Aging Trends Study}},
  title        = {{NHATS-CMS} Linked Data Application Instructions and Guide},
  howpublished = {\url{https://nhatsannex.org/nhats-cms-linked-data-application-instructions-and-guide}},
  note         = {Accessed October 29, 2026}
}

@book{sun2006statistical,
  title={The statistical analysis of interval-censored failure time data},
  author={Sun, Jianguo},
  year={2006},
  publisher={Springer}
}

@misc{nhatsLML,
  author = {{National Health and Aging Trends Study}},
  title  = {{NHATS} Round 2: Last Month of Life Interview, Place of Death},
  note   = {Section PD, Questions PD2A--PD2B},
  url    = {https://www.nhats.org/sites/default/files/public-documentation/NHATS%20Round%203%20LML%20Data%20Collection%20Instrument_0.pdf}
}

@misc{resdac2022death,
  author       = {Jarosek, Stephanie},
  title        = {Death Information in the Research Identifiable Medicare Data},
  howpublished = {Research Data Assistance Center (ResDAC)},
  year         = {2022},
  url          = {https://resdac.org/articles/death-information-research-identifiable-medicare-data}
}

@article{binder1992fitting,
  title={Fitting Cox's proportional hazards models from survey data},
  author={Binder, David A},
  journal={Biometrika},
  volume={79},
  number={1},
  pages={139--147},
  year={1992},
  publisher={Oxford University Press}
}

@inproceedings{jing2019survey,
  author    = {Jing, Bocheng and Diaz-Ramirez, L. Grisell and Boscardin, W. John and Lee, Alexandra K. and Lee, Sei J.},
  title     = {Survival Estimates Prediction with Survey Data},
  booktitle = {Proceedings of SAS Global Forum 2019},
  pages  = {1--12},
  year      = {2019},
  publisher = {SAS Institute Inc.},
  url       = {https://support.sas.com/resources/papers/proceedings19/3900-2019.pdf}
}

@misc{sasSurveyphregSupport,
  author       = {{SAS Institute Inc.}},
  title        = {How Can I Obtain Survival Estimates from {PROC SURVEYPHREG}?},
  howpublished = {SAS Customer Support Knowledge Base, Article KB0057487},
  url          = {https://sas.service-now.com/csm/en/how-can-i-obtain-survival-estimates-from-proc-surveyphreg?id=kb_article_view&sysparm_article=KB0057487},
  note         = {Accessed November 01, 2026}
}

\end{singlespace}

\end{document}